\documentclass[12pt,journal,compsoc]{IEEEtran}
\ifCLASSINFOpdf
\else
\fi
\usepackage[dvips]{graphicx}
\usepackage{cite}
\ifCLASSOPTIONcompsoc
\else
\fi
\ifCLASSINFOpdf
\else
\fi

\begin{document}
\title{Rotational sensitivity  of the ``G-Pisa'' gyrolaser}
\author{Jacopo~Belfi, Nicol\`o~Beverini, Filippo~Bosi,
Giorgio~Carelli, Angela~Di~Virgilio, Enrico~Maccioni, Marco~Pizzocaro
\IEEEcompsocitemizethanks{\IEEEcompsocthanksitem J.~Belfi, N.~Beverini
 G.~Carelli, E.~Maccioni and M.~Pizzocaro are with the Department of Physics ``Enrico Fermi'', Universit\`a di Pisa,
and CNISM unit\`a di Pisa.
E-mail: carelli@df.unipi.it}\protect\\
\IEEEcompsocitemizethanks{\IEEEcompsocthanksitem F.~Bosi and A.~Di~Virgilio
 are with INFN sezione di Pisa.\protect\\
}}

\ifCLASSOPTIONpeerreview
\markboth{Journal of \LaTeX\ Class Files,~Vol.~6, No.~1, January~2007}%
{Shell \MakeLowercase{\textit{et al.}}: Bare Advanced Demo of IEEEtran.cls for Journals}
\fi

\IEEEcompsoctitleabstractindextext{
\begin{abstract}
G-Pisa is an experiment investigating the possibility to operate a high sensitivity laser gyroscope 
with area less than $1\, \rm m^2$ for improving the performances 
of the mirrors suspensions of the gravitational wave antenna Virgo. 
The experimental set-up consists in a He-Ne ring laser with a 4~mirrors square cavity.
The laser is pumped by an RF discharge where the RF oscillator includes the
laser plasma in order to reach a better stability.
The contrast of the Sagnac fringes is typically above 50\%
and a stable regime has been reached with the laser operating both 
single mode or multimode.  The effect of hydrogen contamination on the
laser was also checked.
A low-frequency sensitivity, below $1\,\rm Hz$, in the range of
$10^{-8}\,\rm {(rad / s)/ \sqrt{Hz}}$ has been measured.
\end{abstract}}
\maketitle
\IEEEdisplaynotcompsoctitleabstractindextext
 \ifCLASSOPTIONpeerreview
 \begin{center} \bfseries EDICS Category: 3-BBND \end{center}
 \fi

\IEEEpeerreviewmaketitle

\section{Introduction}
\IEEEPARstart{L}{aser} gyroscopes are devices sensitive to inertial angular motion. 
They are based on the Sagnac effect:
in a closed cavity rotating at angular velocity $\Omega$ the
two counter propagating beams complete the path at different times.
Different kinds of such devices have been developed mainly for navigation.
They are only sensitive to angular velocity and insensitive to translational velocity.
We distinguish between passive (fiber optic gyros) and active (ring lasers) Sagnac interferometers.
Passive devices measure the phase shift between the two beams, while the active
ones measure the frequency difference, an inherently more accurate measurement.
Small fiber gyros are typically used for navigation and have a resolution of $10^{-8}$~rad/s,
while the large ring laser gyros used in geophysics and geodesy reach the level of $10^{-12}$~rad/s.
In the following we will focus on active ring lasers and will call them simply gyros.
One application of large gyros is the monitoring of the variations of the Earth angular velocity vector.
The orientation with respect to the Earth axis is important since the induced signal is proportional to the scalar product between
the normal of the gyro area and Earth axis, see equation 1.
For horizontal gyros the signal is zero at the equator and maximum at the pole;
at intermediate latitudes, horizontal and vertical cavities work fine.
Up to now the reached resolution, integrating the signal for several hours, is $10^{-8}$ of the Earth rotation rate
\cite{{sted97},{schr03},{schr04},{rowe99},{dunn02}}.
The Sagnac frequency, i.e., the beat signal between the two output beams, is:
\begin{equation}\label{eq1}
\delta \phi = 4 A \, \mathbf{n}\cdot \mathbf{\Omega}  / (\lambda P) + \phi _\rho
\end{equation}
 
where $A$ and $P$ are the area and the perimeter of the cavity respectively,
$\lambda$ is the wavelength of the laser beam, $\mathbf{n}$ is the normal vector of the plane 
of the ring cavity and $\mathbf{\Omega}$ is the induced vector of rotation. $\phi _\rho$ is denoting additional, 
usually very small, contributions to the Sagnac frequency 
due to non-reciprocal effects in the laser cavity, such as Fresnel drag \cite{arom03}.
A laser gyro can monitor with high accuracy the orientation of the laboratory reference frame,
but our main interest is in extending its use for improvements of Virgo, the gravitational
waves interferometric antenna \cite{virg08, virg09}.
The major draw-back in ring lasers is the lock-in between the two counter-propagating modes; 
in large gyros the rate bias induced by the Earth rotation is enough to avoid it. 
The large ring laser gyros built by the joint ring laser working group in
New Zealand and Germany have sides of the order of several meters.
``G'' \cite{{schr01},{schr02}} located at the Geodetic Observatory in Wettzell (Germany)
is a monolithic square with 4 m sides, ``UG2'', a rectangle with 40 m by 20 m sides, is the largest ring.
In principle the larger the area the better the sensitivity,
but the best gyro so far, is ``G'', which is operating very close to the quantum limit (just a factor 3 higher),
and its measured noise power spectrum is of the order of $10^{-11}$~rad/s/$\sqrt{\rm Hz}$, with a duty cycle close to 100\%.
The Allan deviation has not shown any systematic effects on a time scale of about 3~hours,
and the sensor drift is normally below 1.5 parts in $10^8$ of the Earth rotation ($1.1 \cdot 10^{-12}$~rad/s).
The Allan deviation goes down as the square root of time up to 2.7~hours of averaging time,
then goes up for a while before it resumes the theoretically expected downward trend.

\section{Sensitivity limits}

\begin{figure}
\centering
\includegraphics[width=3.5in]{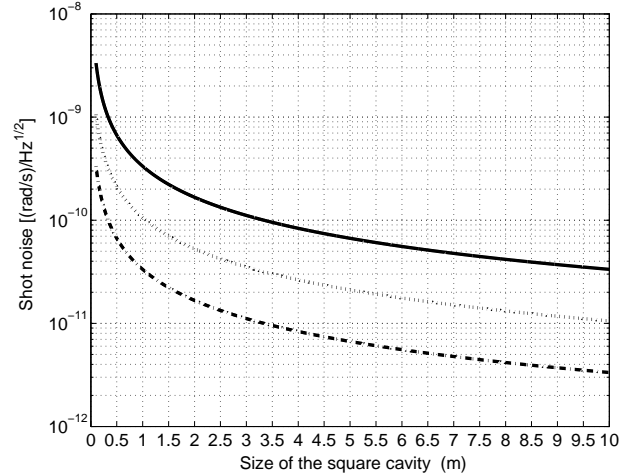}
\caption{Shot noise limit as a function of the square cavity side. The three different traces correspond to sets of mirrors of different quality (see the text).}
\label{shot}
\end{figure}

In large gyros the shot noise of light gives the fundamental limit to sensitivity:

\begin{equation}
 \Omega _{sn} = \frac {c}{\, 2 \pi \,K\,L\,}\, \sqrt {h\,\nu\,\mu
\,\frac {\,T\,}{2P\,t}}
\end{equation}

where $c$ is the speed of light, $L$ the side of the square ring, $h$ the Plank Constant,
$\nu$ the frequency of the light, $\mu$ the total cavity losses, $T$ the transmittance,
$P$ the total transmitted power, $t$ the observation time, and $K$ is the scale factor of the instrument.
In figure \ref{shot} the shot noise is shown for various values 
of the parameters $\mu$, $T$ and $P$ and an observation time of 1~s.

\begin{figure}
\centering
\includegraphics[width=3.5 in]{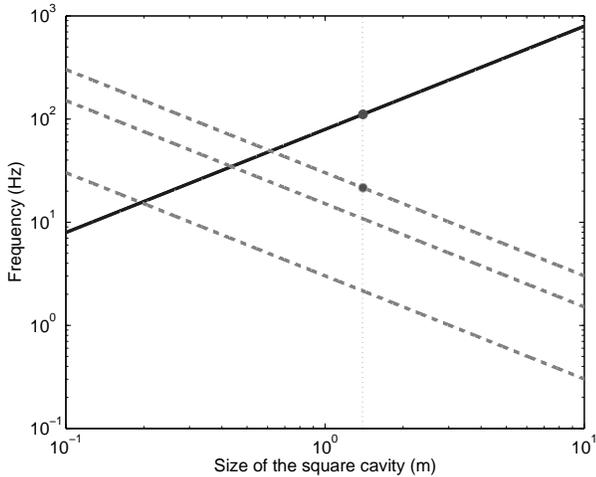}
\caption{Continuous  line shows the Sagnac frequency, the three dashed lines show
the lock-in threshold frequency for three different set
of mirrors (s=10$^{-4}$, 5$\cdot$10$^{-4}$ and 10$^{-3}$). Highlighted intersections represent the G-Pisa case.}
\label{sagnac}
\end{figure}

For the continuous line (good mirrors, but not the best ones) $\mu = 40$~ppm, $T = 0.2$~ppm and $P = 10^{-8}$~W, for the dotted line
absorption has been reduced to 4~ppm, and for the point-dotted line the transmitted power has also been increased
to $10^{-7}$~W (top quality mirrors). In short, a device with $L$ below 1~m could reasonably reach a sensitivity of $10^{-9}$~rad/s/$\sqrt{\rm Hz}$,
while to reach $10^{-11}$~rad/s/$\sqrt{\rm Hz\,}$ devices larger than 2~m are required.
The lower limits to the size of a gyrolaser are given by the backscattering-induced frequency pulling and lock-in.
As a rule, for a given set of mirrors, it is always possible to build a ring large enough that the bias induced
by the Earth rotation is large enough to avoid lock-in. 
The parameter which sets the magnitude of the mode pulling is the lock-in threshold frequency $l$,
where $l \approx c\,s\,\lambda\,/(\,\pi\,d\,P\,)$, where $c$ is the speed of light, $s$ is the fraction of the laser field amplitude which is scattered by each mirror, $\lambda$ is the laser wavelength, $d$ is the beam waist and $P$ is the ring perimeter.
In figure \ref{sagnac} the dashed lines show the lock-in limit as a function of the side of a square ring,
for a beam waist of 0.5~mm and different values of of the scattering coefficient $s$. The continuous line shows the Sagnac frequency
given by the Earth rotation for a device horizontally located at latitude 43$^\circ$ as a function of the ring size.

\section{Experimental set-up}
G-Pisa is a square cavity 5.60~m in perimeter and 1.96~m$^2$ in area. The mechanical  design is flexible and each side of the square 
can be scaled from 1.40~m down to 0.90~m  with minor changes.
The mechanical system is mounted onto an optical table and  has a stainless steel modular structure: 4 boxes, located at the
corner of the square and containing the mirrors holders inside, are connected by tubes, so to form a ring vacuum chamber with a 
total volume of about $5\cdot 10^{-3}$~m$^3$. The vacuum chamber, is entirely filled with a mixture
of He and a 50\% isotopic mixture of $^{20}$Ne and $^{22}$Ne. The total pressure of the gas mixture is set to 560~Pa with
 a partial pressure of Neon of 20~Pa. In the center of one of the tubes there is a pyrex insertion, a
capillary with 4 mm internal diameter, approximately 15 cm long, which is the discharge tube.
While the discharge in the laser medium of ``G'' and analog systems is excited by an RF source,
where two coils couple the RF oscillator to the gas, in our system we choose a
capacitive coupling. Two halves of a copper cylinder are used 
as electrodes and are part of the resonant circuit of the RF source.
The laser medium in this way is included in the active circuit and the fluctuations in the plasma
density do not affect the coupling of the RF source to the discharge, but only the oscillator
frequency, about 115~MHz.
This kind of discharge provides a very good passive stability and made possible to
regulate the laser output power very close to the laser threshold since the first runs.
The discharge is 5~cm long, but different lengths will be tested in the next months 
to minimize the effect of plasma intensity fluctuations around the laser threshold.
The discharge tube has four micro-metric screws used to align the pyrex capillary with the mirrors
and the optical cavity. 

Four spherical mirrors with 6~m radius curvature were chosen for the resonator, and two micro-metric
lever arms acting on the tilts of each mirror, make it possible a fine tuning of the cavity alignment.
Mirrors reflectivity is optimized for the emission line around 632.8~nm.
The free spectral range of the cavity is 53.6~MHz, the horizontal beam waist is 0.68~mm,
the sagittal beam waist 0.56~mm and the intra-cavity ring-down time is 20~$\mu$s. 
In order to achieve  long term stability of the perimeter the laser gyro optical frequency will be locked to a reference laser. 
This will involve the measurement of the radio frequency beat note between the gyrolaser output and the reference laser: 
for such application the capacitive coupling will introduce less noise than the inductive one.
The two outputs (clockwise and counter-clockwise sense of circulation) from one cavity
mirror are combined by means of a 50\% intensity beam splitter and detected by a photodiode.
The photodiode current is voltage converted by a trans-impedance stage with a gain of $10^9$
 and  1~ms rise time. The two single beam outputs are also monitored by means of two fiber-coupled photomultipliers. The laser modal structure has been detected injecting the output beams in a high finesse linear cavity.
The signals are acquired and analyzed off-line.

\section{Experimental Results}
\begin{figure}
\centering
\includegraphics[width=3.5in]{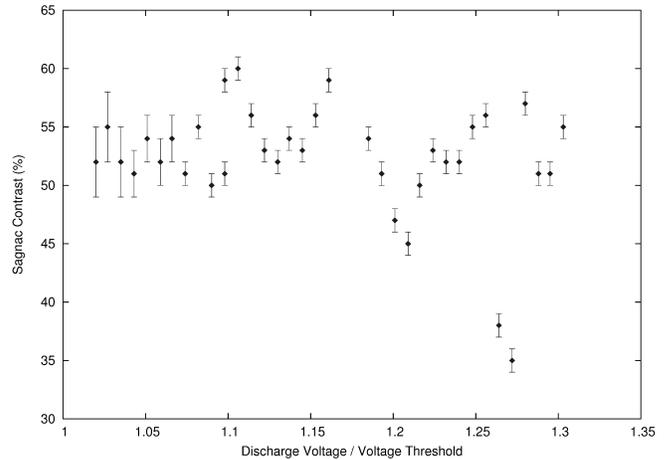}
\caption{Contrast of the Sagnac fringes in function of the discharge root mean square voltage normalized to the threshold value.}
\label{contrast}
\end{figure}

When the laser output  power is not higher than few tens~nW only one longitudinal mode
(TEM$_{00}$) in both sense of propagation is above the laser threshold.
The beat note at the expected Sagnac frequency (around 111.1~Hz) can be observed both in the
interference between the two counter-rotating beams and in the single beam.
Usually both beams lase in the same longitudinal mode, i.e., with the
same mode number. This behavior is the standard operation for a ring laser gyro, and the beat
note between the two counter-rotating waves is the Sagnac frequency at the expected value.
Unfortunately, counter-propagating beams can lase on different longitudinal
modes and one gets the so called split-mode regime \cite{Hurst}. Their separation is 53.6~MHz, the free spectral range of the gyro laser.
Sagnac signal is lost with split modes. The Sagnac signal in the beat note between the two
split counter-propagating modes still exists, of course, but is shifted around 53.6~MHz. 
The first cause of split in G-Pisa is mode hopping due to thermal expansion of the cavity. In the not controlled
conditions of our laboratory, the typical time between one mode
jump and the other could range between 5 min to 15 min. Split modes persist
usually for some minutes, until both beams return on the same longitudinal
mode and then the standard operation and Sagnac signal are recovered.
When the gyrolaser operates close to threshold, as stated before, just one optical mode for each direction propagates inside the cavity,
if the power increases slightly we obtain multimode operation, at higher power  many different longitudinal modes are 
excited in both directions and the behavior of the gyroscope becomes unstable.
Several measurements were done to learn about the  behavior of the ring laser as a function
of power and to investigate multimodal operation of the cavity.
During the standard operation, G-Pisa gyro shows a stable Sagnac signal, even with three or four modes in each direction. This
evidence strongly suggests that modes are apparently mode-locked, or, there
is a fixed phase relation between different modes, and this fixed phase is the
same in both directions. Contrast of the Sagnac signal seems uncorrelated to intensity of the laser
beam or number of modes. Figure \ref{contrast}  shows an example, with number of
modes ranging between one to four (fluctuations are probably due to hysteresis effects and are not strictly reproducible). 
From 1 up to 1.1 normalized voltage it has been checked that the laser operates in monomode.
We noted that the gyro sometimes does not mode lock. This happens especially after power changes and for particular distributions 
of the cavity modes under the gain curve. These last aspects deserve further investigation in the next future.



\begin{figure}
\centering
\includegraphics[width=3.5 in]{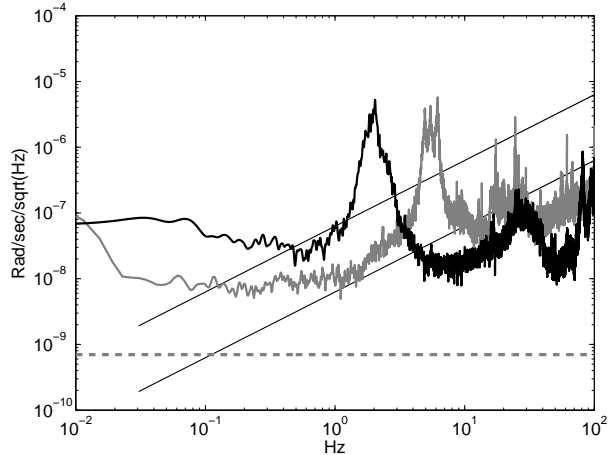}
\caption{Comparison between the best low-frequency measurements. Gray spectrum and black spectrum refer respectively to the case of rigid 
and passively air-damped optical  table. The horizontal line is the shot noise, the transverse 
lines represent the minimal and typical request for Virgo.}
\label{noise}
\end{figure}

The rotational sensitivity is obtained by reconstructing the phase of the beat note,
and differentiating it, in order to obtain the angular velocity.
The mean value of this signal gives the Earth rotational speed, which is usually subtracted.
The power spectrum of the signal gives the upper limit to the low frequency sensitivity,
which is the relevant parameter for the possible improvements of the gravitational waves interferometer suspension.
A typical spectrum is reported in fig. \ref{noise}. The horizontal line is the shot noise limit, 
the two transverse lines are the minimal and typical requirements for the Virgo suspensions \cite{brac05}.

\begin{figure}
\centering
\includegraphics[width=3.5in]{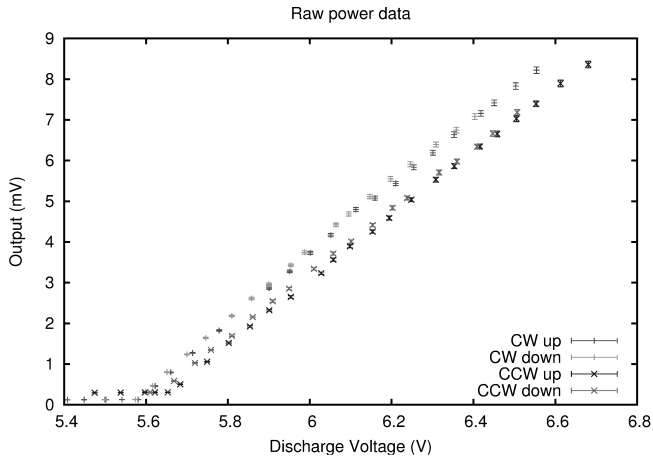}
\caption{Clockwise and counter-clockwise beams in function of the voltage applied to the discharge}
\label{power}
\end{figure}

Fig. \ref{power} shows the behavior of the intensity of clock wise and anti-clock wise beams,
measured increasing (up) and decreasing (down) the voltage of the discharge.

\begin{figure}
\centering
\includegraphics[width=3.5in]{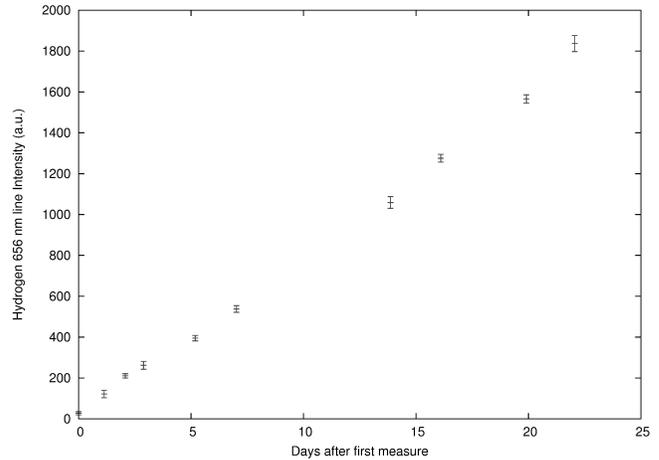}
\caption{Hydrogen contamination as a function of time}
\label{hydrogen}
\end{figure}

It is possible to see that the two modes have different threshold and
different slopes. This put in evidence a slight intrinsic asymmetry in the laser.
By means of a simultaneous measurement of rotational noise, performed with the G-Pisa gyrolaser, and translational noise obtained by a 
three axial linear accelerometer, it was possible to set an upper limit to the correlation between detected rotations
 and translations to a level of 0.5$\%$. Details of such analysis can be found in \cite{virg08}.
Hydrogen contamination is a matter of concern in our apparatus, since light absorption from hydrogen
can prevent laser action. Hydrogen is released from the steel vacuum tubes and from the pyrex 
close to the discharge.
Fig.~\ref{hydrogen} shows the concentration of Hydrogen as a function of time.
G-Pisa can now run for only three weeks before the gas mixture must be changed.
This level of contamination will be reduced by introducing some getters inside the chamber and by baking
the vacuum chamber before each gas refilling.

\section{Conclusions}

Our device can run continuously, interruptions are due to thermal expansion induced mode jumps (split mode regime), which makes the instrument
blind for few minutes. This problem will be resolved by stabilizing the perimeter of the ring by means of piezoelectric actuated mirror.
We characterized the different regimes of operation of the gyroscope and proved the possibility of increased power, multimode operations.
The low frequency measured power spectrum (below 1 Hz) is around $10^{-8}\,\rm (rad/s)/\sqrt{Hz\,}$, which looks like
to be real motion. The system sensitivity is close to the limit needed to control the tilt
of the suspensions of Virgo.


\begin{thebibliography}{1}

\bibitem{sted97}
G. E. Stedman, ``Ring-laser tests of fundamental physics and geophysics,''
\textit{Rep. Prog. Phys.}, vol. 60, pages 615--688, 1997.

\bibitem{schr03}
K. U. Schreiber, T. Kl\" ugel and G. E. Stedman,`` Earth tide and tilt detection by a ring laser gyroscope,'' 
\textit{J. Geophys. Res. Solid Earth}, vol. 108, 2003.

\bibitem{schr04}
K. U. Schreiber, A. Velikoseltsev, M. Rothacher. T. Kl\" ugel, G. E. Stedman and D. Wiltshire,
``Direct measurement of diurnal polar motion by ring laser gyroscopes,'' 
\textit{J. Geophys. Res. Solid Earth}, vol. 109, B06405, 2004.

\bibitem{rowe99}
C.H. Rowe, U.K. Schreiber, S.J. Cooper, B.T. King, M. Poulton and G.E. Stedman, 
``Design and operation of a very large ring laser gyroscope,'' 
\textit{Appl. Opt.}, vol. 38, pages 2516--2523, 1999.

\bibitem{dunn02}
R.W. Dunn, D.E. Shabalin, R.J. Thirkettle, G.J. MacDonald, G.E. Stedman and K.U. Schreiber,
 ``Design and Initial Operation of a 367 $\mathrm{m}^2$ Rectangular Ring Laser,'' 
\textit{ Appl. Opt.}, vol. 41 (9), pages 1685--1688, 2002.

\bibitem{arom03}
F. Aronowitz, in ``Laser applications``, vol. 1 M. Ross ed., Academic Press, New York 1971. 



\bibitem{virg08}
A. Di Virgilio et al., ``The G-Pisa gyrolaser after 1 year of operation and considerations 
about its use to improve the Virgo IP control'', Virgo note (to be published).

\bibitem{virg09}
I. Fiori, S. Braccini, F. Travasso and A. Vicer\'e  ``Siesta simulation of NE suspension tower Virgo note'',
VIR-NOT-PIS-1390-285, 2004.

\bibitem{schr01}
K.U. Schreiber, A Velikoseltsev, Kl\"ugel T., G.E. Stedman and W. Schl\"uter, 
``Advances in the Stabilisation of Large Ring Laser Gyroscopes,'' 
\textit{Proceedings of the Symposium Gyro Technology}, Univ. of Stuttgart, Stuttgart, Germany, 2001. 

\bibitem{schr02}
K. U. Schreiber, T. Kl\"ugel, G. E. Stedman, and W. Schl\"uter,
``Stabilitaetsbetrachtungen f\"ur grossereinglaser,'' DGK Mitteilungen Rehie A., Heft 118, pages 156--158, 2002. 

\bibitem{Hurst} R. B. Hurst, R. W. Dunn, K. U. Schreiber, R. J. Thirkettle, and G. K.
MacDonald, `` Mode behavior in ultralarge ring lasers,'' Appl. Opt. 43,
2337, 2004.


\bibitem{brac05}
S. Braccini et al.`` Measurement of the seismic attenuation performance of the VIRGO Superattenuator'', \textit{Astrop. Phys.},vol. 23, 
pages 557--565, 2005.


\end{thebibliography}
\end{document}